\documentclass[acus]{JAC2000}

\usepackage{graphicx}

\begin{document}
\title{RESULTS ON PLASMA FOCUSING OF HIGH ENERGY DENSITY 
ELECTRON AND POSITRON BEAMS \thanks{Work supported in part by the 
Department of Energy under contracts 
DE-AC02-76CH03000, DE-AC03-76SF00515, DE-FG03-92ER40695,
and DE-FG05-91ER40627, and the Univ. of California
Lawrence Livermore National Laboratory, 
through the Institute for Laser Science and Applications,
under contract No. W-7405-Eng-48; and by the 
US-Japan Program for Cooperation in High Energy Physics.}}

\author{J.S.T.~Ng, P.~Chen, W.~Craddock, F.J.~Decker, R.C.~Field, 
M.J.~Hogan, R.~Iverson,\\
F.~King, R.E.~Kirby, T.~Kotseroglou,
P.~Raimondi, D.~Walz, SLAC, Stanford, CA. 94309, USA\\
H.A.~Baldis\thanks{Also at UC Davis, Dept. of Applied Science.}, 
P.~Bolton, LLNL, Livermore, CA. 94551, USA\\
D.~Cline, Y.~Fukui, V.~Kumar, UCLA, Los Angeles, CA. 90024, USA\\
C.~Crawford, R.~Noble, FNAL, Batavia, IL. 60510, USA \\
K.~Nakajima,  KEK, Tsukuba, Ibaraki 305-0801, Japan\\
A.~Ogata, Hiroshima University, Kagamiyama, Higashi-Hiroshima, 739-8526 Japan\\
A.W.~Weidemann, University of Tennessee,  Knoxville, Tennessee 37996, USA\\
}

% R.~Alley$^{1}$, 
% P.~Anthony$^{1}$,
% S.~Chattopadhyay$^{3}$,
% D.D.~Meyerhofer$^{6}$,
% A.~Sessler$^{3}$,
%\address{
%\vspace{10pt}
% { \renewcommand{\baselinestretch}{1.0} \Large \normalsize
% $^{3}${\it Lawrence Berkeley National Laboratory, Berkeley, California}\\
% $^{6}${\it Department of Physics and Astronomy, University of Rochester,
% Rochester, New York 14627} \\
%
% }
%}

%\date{\today}
\maketitle

\begin{abstract} 

We present results from the SLAC E-150 experiment on plasma focusing of 
high energy density electron and, for the first time, positron beams.
We also discuss measurements on plasma lens-induced synchrotron radiation, 
longitudinal dynamics of plasma focusing, and
laser- and beam-plasma interactions.

\end{abstract}

\section*{Introduction}
The plasma lens was proposed as a final focusing mechanism to achieve high 
luminosity for future high energy linear colliders \cite{chen}. 
Previous experiments to test this concept were carried out with low 
energy density electron beams \cite{ucla-kek-lbl}.
In this paper, we present preliminary results obtained recently by the E-150
collaboration on plasma focusing of high energy density electron and
positron beams.

\begin{table}[htb]
\caption{FFTB electron and positron beam parameters for this experiment.}
\label{fftb:parm}
%\begin{tabular}{lrrr}
%\begin{tabular}{lccc}
\begin{tabular}{lll}

Parameter       & Value    \\ \hline
Bunch intensity & $1.5 \times 10^{10}$ particles per pulse \\
Beam size       & 5 to 8~$\mu$m (X), \\
                & 3 to 5~$\mu$m (Y) \\
Bunch length    & 0.7~mm  \\
Beam energy     & 29~GeV \\
Normalized emittance       
                & 3 to 5 $\times 10^{-5}$ m-rad (X), \\
                &  0.3 to 0.6 $\times 10^{-5}$ m-rad (Y)\\
Beam density    & $\sim 7 \times 10^{16}$ cm$^{-3}$ \\  \hline
\end{tabular}
\end{table}

\section*{Experimental setup}

The experiment was carried out at the SLAC Final Focus Test Beam 
facility (FFTB)\cite {fftb_ref}.  The experiment operated parasitically with
the PEP-II B-factory; the high energy electron and positron beams
were delivered to the FFTB at 1 - 10 Hz from the SLAC linac.
The beam parameters are summarized in Table~\ref{fftb:parm}.  

\begin{figure*}[t]
\centering
\includegraphics*[width=165mm]{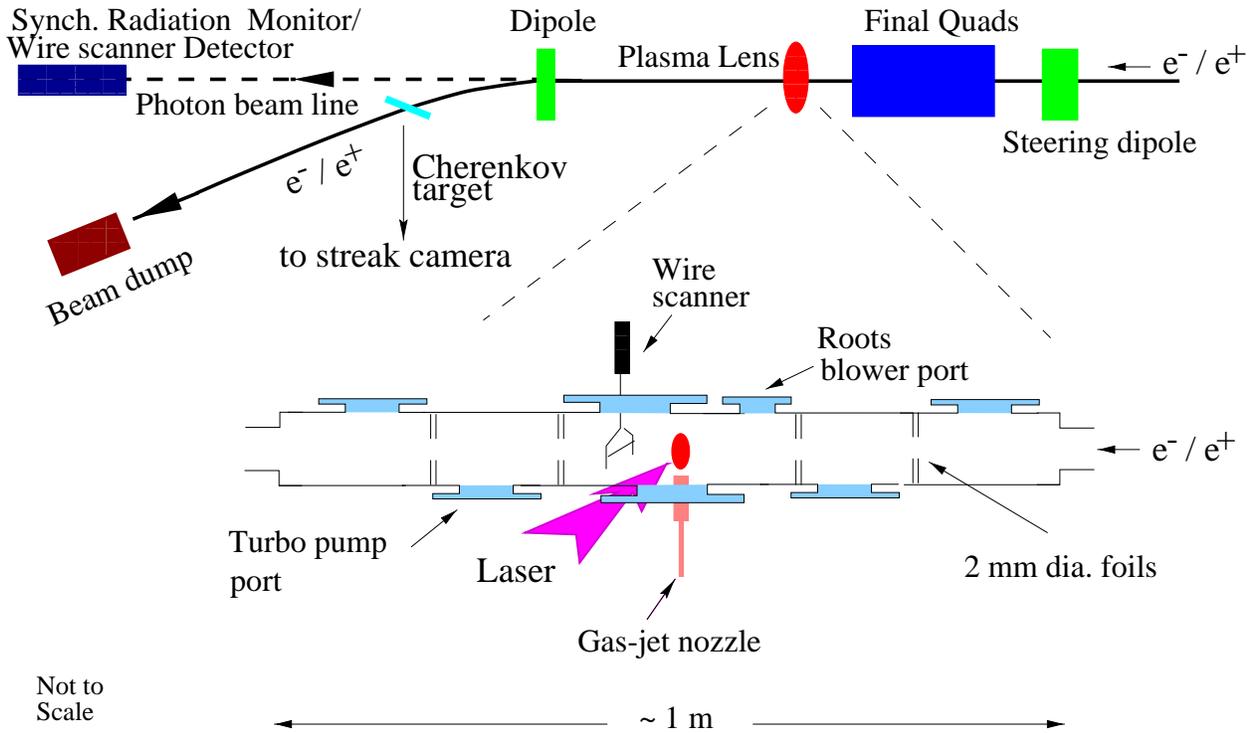}
\caption{Layout of the plasma lens measurement setup and schematics of
the plasma chamber.}
\label{beamline}
\end{figure*}

A layout of the beam line and a schematic drawing of the plasma chamber
are shown in Figure~\ref{beamline}.  
The beam size was measured using a wire scanner system.
A carbon fiber $4~\mu$m or $7~\mu$m
in diameter was placed downstream of the plasma lens, adjustable 
along the beam axis in a range of 8 to 30 mm from the center of the lens.  
The Bremsstrahlung photons were detected in a Cherenkov type detector 
located 35~m downstream of the lens.
A set of ionization chambers interleaved with polyethylene blocks,
located 33~m downstream of the lens, was used to monitor the
synchrotron radiation emitted as a result of the strong bending of
the beam particles by the plasma lens.  This detector provided an
independent measure of the focusing strength. 
Also, a Cherenkov target was installed
in the electron beam line downstream to enable streak camera diagnostics
of the longitudinal plasma focusing dynamics.

To create the plasma lens, a short burst (800~$\mu$s duration) 
of neutral nitrogen or hydrogen
gas, injected into the plasma chamber by a fast-pulsing nozzle,
was ionized by a laser and/or the high energy beam.
The neutral density was determined by interferometry to
be $4 \times 10^{18}~\rm{cm}^{-3}$ for N$_2$ and 
$5 \times 10^{18}~\rm{cm}^{-3}$ for H$_2$ at a plenum pressure of 1000~psi.
The injected gas was evacuated by a differential pumping system which
made operation of the gas jet possible while maintaining ultra-high vacuum in
the beam lines on either side of the chamber.  

\section*{Plasma focusing}
For a bunched relativistic beam traveling in vacuum, the Lorentz force
induced by the collective electric and magnetic fields is nearly cancelled,
making it possible to propagate over kilometers without significant
increase in its emittance.
In response to the intruding beam charge and current, the
plasma electron distribution is re-configured to neutralize the 
space charge of the beam and thereby cancel its radial electric field.  
For a positron beam, the plasma electrons are
attracted into the beam volume thus neutralizing it; for an electron beam, the
plasma electrons are expelled from the beam volume, leaving behind the
less mobile positive ions which neutralize the beam.  When the beam radius 
is much smaller than the plasma wavelength, the neutralization of the
intruding beam current by the plasma return current is ineffective
because of the small skin depth.  This leaves the azimuthal magnetic
field unbalanced which then ``pinches'' the beam.
In this experiment, typical plasma densities were of the order of
$10^{18}~{\rm cm}^{-3}$, corresponding to a plasma wavelength of
approximately 30~$\mu$m which was indeed 
much larger than the incoming beam radius.

The plasma was created by means of beam self-induced ionization and
laser avalanche ionization.
For the case of beam self-ionization,
a small fraction of the neutral gas molecules was ionized due to
collisions with the high energy beam particles.
The secondary electrons from this impact ionization process were 
accelerated by the intense collective field in the beam, transverse to
the direction of propagation, to further ionize the gas \cite{simon_yu}.
That is, the head of the bunch was able to ionize the gas while the
core and the tail of the bunch were focused.  A more quantitative 
understanding requires detailed calculations which are not yet 
available for this experimental setup. 

The results on laser pre-ionization plasma focusing
were obtained using a turn-key infrared ($\lambda = 1064$~nm) laser 
system.  It delivered 1.5~Joules of energy per pulse of 10~ns FWHM
at 10~Hz.  The laser light was
brought to a line focus at the gas jet; the plasma thus produced was
approximately 0.5~mm thick as seen by the $e^{+}/e^{-}$ beams.

With the relatively long infrared laser pulse, the pulse front
was able to ionize a small fraction of the gas by multiple-photon absorption;
the resulting secondary electrons were accelerated, transverse to 
the laser's incident direction, to further ionize the gas.
This process led to an avalanche growth in plasma density, similar to
the beam self-ionization case.
\subsection{Results on plasma focusing}
\begin{figure}[t]
\centering
\includegraphics*[width=80mm]{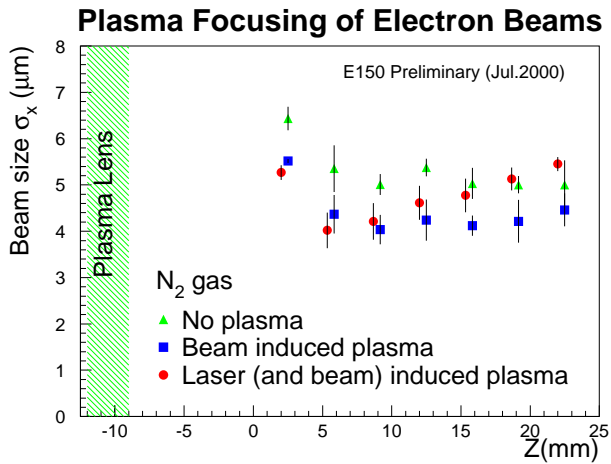}
\includegraphics*[width=80mm]{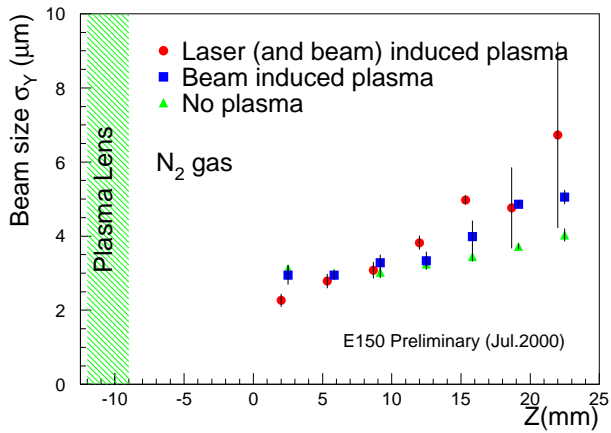}
\caption{Plasma focusing for electron beams in the X (top) and Y (bottom)
dimensions.}
\label{electrons}
\end{figure}

\begin{figure}[t]
\centering
\includegraphics*[width=80mm]{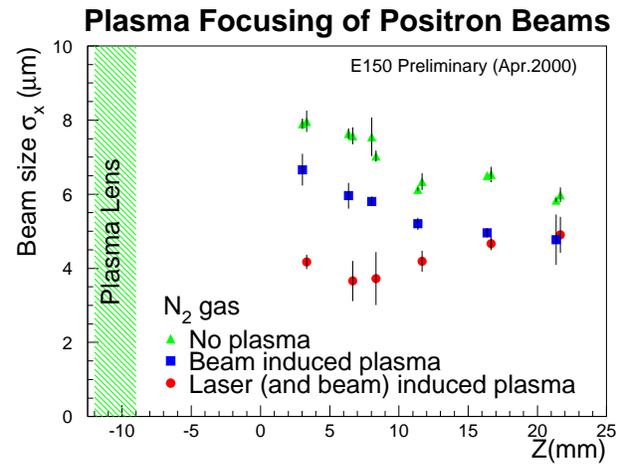}
\includegraphics*[width=80mm]{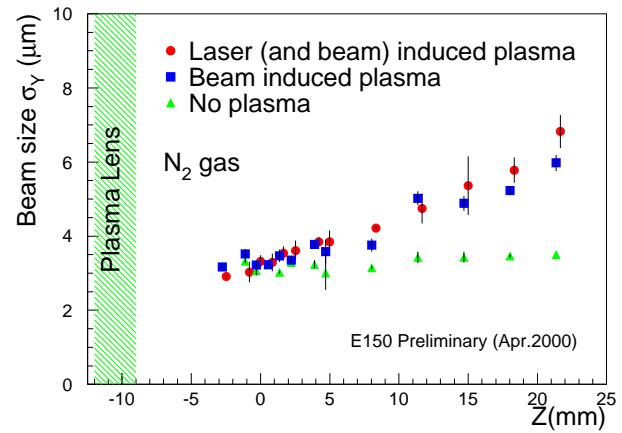}
\caption{Plasma focusing for positron beams in the X (top) and Y (bottom)
dimensions.}
\label{positrons}
\end{figure}

The results for laser (and beam) 
ionization plasma focusing of electron and positron beams
are shown in Figures~\ref{electrons} and \ref{positrons}, respectively.  
The measured transverse beam size is shown as a function of the 
distance (Z) between the wire scanner and the
plasma lens.  The axis of the gas jet is at Z = -10.5 mm.
In the X-dimension, the beam envelope is shown converging
without plasma focusing (triangle points); while
with laser (and beam) induced plasma focusing (filled circles),
the beam envelope is shown converging towards a reduced waist and then
diverging because of the strong focusing.
In the Y-dimension, the waist is at a location close to the 
the plasma lens beyond the reach of the
wire scanner; the beam envelope is seen diverging due to
the strong plasma focusing.  Focusing is also observed for
beam-induced plasma with the laser turned off.

\section*{Other results}
Discussions on additional results obtained from this experiment can be
found in \cite{aac}.  A brief summary is given here.

During the plasma focusing measurements, the focusing strength was 
also measured independently by monitoring the synchrotron radiation 
emitted by particles focused by the lens. The critical energy was
estimated to be a few MeV, corresponding to a focusing gradient of
$10^6$~T/m.  The longitudinal focusing
dynamics was diagnosed with a streak camera with pico-second time
resolution and, as expected, the focusing
was strongest at the longitudinal center of the bunch.  The 
laser- and beam-plasma interaction was studied by varying the laser
pre-ionization timing with respect to the beam arrival time; we 
observed a delay-correlated modulation of the plasma focusing in the
``after-glow'' regime.

\section*{SUMMARY AND OUTLOOK}
Results on plasma focusing of 29 GeV electron and, for the first time,
positron beams have been presented.  Beam self-ionization turned out to be
an economical method for producing a plasma lens.
The infrared laser with a
10 ns long pulse also proved to be efficient in plasma production, resulting
in the strong focusing of electron and positron beams.  
Data on other aspects of plasma focusing were also collected; detailed
discussion is presented elsewhere \cite{aac}.

Design studies for linear collider applications are just starting.
The first issue to resolve is the effect of beam jitter on the
achievable luminosity of plasma focused beams.  Optimization of 
plasma lens parameters requires bench-marking
of computer codes as well as better understanding of the various plasma
production processes.  The experience gained in this experiment
will serve as a basis for further engineering design studies for
an eventual plasma lens application.

\end{document}